\documentclass[]{aa} 
\usepackage{epsfig,graphicx,threeparttable,times,multirow,multicol,amsmath,amssymb,color,longtable,breqn}
\usepackage[T1]{fontenc}
\usepackage[varg]{txfonts}

\usepackage{natbib,twoopt}
\usepackage[hyphenbreaks]{breakurl}
\usepackage[breaklinks]{hyperref}      
\bibpunct{(}{)}{;}{a}{}{,}             
\definecolor{cobalt}{rgb}{0.06, 0.2, 0.65}
\hypersetup{
  colorlinks,
  citecolor=cobalt,
  linkcolor=cobalt,
  urlcolor=cobalt,
}
\makeatletter
  \newcommandtwoopt{\citeads}[3][][]{\href{http://adsabs.harvard.edu/abs/#3}%
    {\def\hyper@linkstart##1##2{}%
     \let\hyper@linkend\@empty\citealp[#1][#2]{#3}}}
  \newcommandtwoopt{\citepads}[3][][]{\href{http://adsabs.harvard.edu/abs/#3}%
    {\def\hyper@linkstart##1##2{}%
     \let\hyper@linkend\@empty\citep[#1][#2]{#3}}}
  \newcommandtwoopt{\citetads}[3][][]{\href{http://adsabs.harvard.edu/abs/#3}%
    {\def\hyper@linkstart##1##2{}%
     \let\hyper@linkend\@empty\citet[#1][#2]{#3}}}
  \newcommandtwoopt{\citeyearads}[3][][]%
    {\href{http://adsabs.harvard.edu/abs/#3}
    {\def\hyper@linkstart##1##2{}%
     \let\hyper@linkend\@empty\citeyear[#1][#2]{#3}}}
\makeatother

\def\ltsima{$\; \buildrel < \over \sim \;$}
\def\lsim{\lower.5ex\hbox{\ltsima}}
\def\gtsima{$\; \buildrel $\geq$ \over \sim \;$}
\def\gsim{\lower.5ex\hbox{\gtsima}}
\newcommand{\be}{\begin{equation}}
\newcommand{\en}{\end{equation}}

\def\deg {$^\circ$}

\def\flux {\mbox{erg~cm$^{-2}$~s$^{-1}$}}
\def\lum {\mbox{erg~s$^{-1}$}}

\def\uu {\object{4U\,0142$+$614}}
\def\ee {\object{1E\,1048.1$-$5937}}
\def\kes {\object{1E\,1841$-$045}}
\def\aa {\object{1E\,1547.0$-$5408}}

\def\rxs {\object{1RXS\,J170849.0$-$400910}}
\def\xte{\object{XTE\,J1810$-$197}}
\def\wes{\object{CXOU\,J164710.2$-$455216}}
\def\ea{\object{1E\,2259$+$586}}
\def\sgra{\object{SGR\,1806$-$20}}
\def\sgrb{\object{SGR\,1900$+$14}}
\def\sgrd{\object{SGR\,1627$-$41}}

\newcommand\cxo{{\em Chandra}}

\newcommand{\A}{{\em ASCA}}

\newcommand{\ein}{{\em Einstein}}
\newcommand\xmm{{\em XMM--Newton}}
\newcommand{\nustar}{{\em NuSTAR}}

\newcommand{\su}{{\em Suzaku}}

\newcommand{\swift}{{\em Swift}}

\newcommand{\integral}{INTEGRAL}
\newcommand{\nicer}{{\em NICER}}

\begin{document}

\title{The long-term enhanced brightness of the magnetar \aa}

\author{
Francesco~Coti~Zelati\inst{1,2}
\and Alice~Borghese\inst{1,2} 
\and Nanda~Rea\inst{1,2}
\and Daniele~Vigan\`o\inst{3,4,1,2} 
\and Teruaki~Enoto\inst{5}
\and \\ Paolo~Esposito\inst{6,7} 
\and Jos\'e~A.~Pons\inst{8}
\and Sergio~Campana\inst{9} 
\and  Gian~Luca~Israel\inst{10}
}

\institute{Institute of Space Sciences (ICE, CSIC), Campus UAB, Carrer de Can Magrans, S/N, 08193, Barcelona, Spain\\
  \email{cotizelati@ice.csic.es}
 \and
 Institut d'Estudis Espacials de Catalunya (IEEC), Gran Capit\`a 2-4, E--08034 Barcelona, Spain
  \and
Departament de F\'isica, Universitat de les Illes Balears, Palma de Mallorca, Baleares E-07122, Spain 
  \and
Institut Aplicacions Computationals (IAC3), Universitat de les Illes Balears, Palma de Mallorca, Baleares E-07122, Spain 
  \and
Department of Astronomy, Kyoto University, Kitashirakawa-Oiwake-cho, Sakyo-ku, Kyoto 606-8502, Japan
  \and
Scuola Universitaria Superiore IUSS Pavia, piazza della Vittoria 15, 27100 Pavia, Italy
  \and
INAF -- Istituto di Astrofisica Spaziale e Fisica Cosmica di Milano, via A. Corti 12, 20133 Milano, Italy
  \and
Departament de Fisica Aplicada, Universitat d'Alacant, Ap. Correus 99, E-03080 Alacant, Spain
\and
INAF -- Osservatorio Astronomico di Brera, via Emilio Bianchi 46, 23807 Merate (LC), Italy
\and
INAF -- Osservatorio Astronomico di Roma, via Frascati 33, 00040 Monte Porzio Catone (Roma), Italy
}

\date{Received 15 September 2019  / Accepted 21 November 2019}

\abstract{
We present the evolution of the X-ray emission properties of the magnetar \aa\ since February 2004 over a time period covering 
three outbursts. We analyzed new and archival observations taken with the \swift, \nustar, \cxo\ and \xmm\ X-ray satellites. The source has 
been observed at a relatively steady soft X-ray flux of $\approx$\,$10^{-11}$ \flux\ (0.3--10\,keV) over the last 9 years, which is about an 
order of magnitude fainter than the flux at the peak of the last outburst in 2009, but a factor of $\sim$\,30 larger than the level in 2006. The 
broad-band spectrum extracted from two recent  \nustar\ observations in April 2016 and February 2019 showed a faint hard X-ray emission 
up to $\sim$\,70\,keV. Its spectrum is adequately described by a flat power law component, and its flux is $\sim$\,7\,$\times$\,$10^{-12}$ \flux\ 
(10--70\,keV), that is a factor of $\sim$\,20 smaller than at the peak of the 2009 outburst. The hard X-ray spectral shape has flattened significantly 
in time, which is at variance with the overall cooling trend of the soft X-ray component. The pulse profile extracted from these \nustar\ pointings 
displays variability in shape and amplitude with energy (up to $\approx$\,25\,keV). Our analysis shows that the flux of \aa\ is not yet decaying to 
the 2006 level and that the source has been lingering in a stable, high-intensity state for several years. This might suggest that magnetars can 
hop among distinct persistent states that are probably connected to outburst episodes and that their persistent thermal emission can be almost 
entirely powered by the dissipation of currents in the corona.
}


\titlerunning{The prolonged enhanced brightness of \aa}

\authorrunning{F.~Coti~Zelati et al.}

\maketitle

\section{Introduction}
\label{sec:intro}

The current census of the isolated neutron star (NS) population includes 26 magnetars, i.e. NSs endowed with an ultra-strong magnetic field 
(typically $B$\,$\sim$\,$10^{13}$--$10^{15}$\,G), whose dissipation is thought to provide the driver for their emission (see \citealt{turolla15, kaspi17, 
esposito18} for recent reviews). With the notable exception of 1E\,161348$-$5055 at the centre of the supernova remnant RCW~103, which 
holds the record as the slowest magnetar ever observed with a spin period of about 6.67\,h (e.g. \citealt{rea16}), the spin periods of these NSs 
lie in a restricted range of values, $\sim$\,0.3--12\,s. Magnetars show a rich phenomenology of high-energy transient episodes, including X-ray 
and gamma-ray bursts lasting from milliseconds to hundreds of seconds, as well as outbursts in which the X-ray flux rises up by a factor between 
a few and several orders of magnitude, and subsequently decreases again on timescales from weeks to years (see the Magnetar Outburst Online 
Catalog\footnote{\url{http://magnetars.ice.csic.es.}}; \citealt{cotizelati18}).

\aa\ was discovered in 1980 with the \ein\ satellite \citep{lamb81} and proposed as a magnetar only three decades later based on the considerable 
long-term X-ray variability (\citealt{gelfand07}). It was identified conclusively as a magnetar following the discovery of pulsations at a period of 
$\sim$\,2.07\,s in the radio band \citep{camilo07}, later detected also in the X-ray band \citep{halpern08}.

\aa\ was caught at an observed flux of $\approx$\,2$\times$10$^{-12}$\,\flux\ (0.3--10\,keV) both with \ein\ in 1980 and with \A\ in 1998. The following 
pointings with \xmm\ in 2004 and \xmm\ and \cxo\ in 2006 found the source at a flux of $\sim$\,(3-4)\,$\times$\,10$^{-13}$\,\flux\ (0.3--10\,keV) -- the 
minimum measured so far. In June 2007, the \emph{Neil Gehrels Swift Observatory} detected \aa\ at a flux larger than that measured in 2006 by a 
factor of $\sim$\,20. Follow-up observations over the subsequent few months revealed that \aa\ was recovering from an outburst occurred before June 
2007 \citep{halpern08}. The source emitted a magnetar-like burst on 2008 October 3, which marked the onset of a second outburst with an abrupt 
increase in the soft X-ray flux by a factor of $\sim$\,200 above the value measured in 2006 \citep{israel10}. It then experienced a state of extreme 
bursting activity on 2009 January 22 (e.g., \citealt{savchenko10}), which coincided with the most powerful outburst hitherto detected from this source. 
The soft X-ray flux increased by a factor of $\sim$\,260 above the value observed in 2006, up to $\sim$\,$8 \times 10^{-11}$\,\flux\ (0.3--10\,keV; 
\citealt{bernardini11, ng11,scholz11}). Emission was also detected at higher energies, up to at least 200\,keV. The flux was larger in the hard X-rays 
than in the soft X-rays by a factor of $\sim$\,5 during the initial phases of the outburst \citep{enoto10,kuiper12}.

Multiple expanding X-ray rings were detected around the source during the first weeks of the outburst decay. They were interpreted in terms of scattering 
of photons emitted during a bright burst by different layers of interstellar dust (\citealt{tiengo10}; see also \citealt{pintore17}). Spectral analysis of these 
structures led to an estimate for the source distance of about 4--5\,kpc \citep{tiengo10}. In the following, we adopt a value of 4.5\,kpc. The value for the 
dipolar component of the magnetic field at the polar caps, assuming the long-term average value for the spin-down rate ($\dot{P}$\,$\sim$\,$4.77\times10^{-11}$\,s\,s$^{-1}$; 
\citealt{dib12}), is $B_{\rm{p}}$\,$\sim$\,$6.4 \times 10^{19} (P\dot{P})^{1/2}$\,G $\sim$\,$6.4\times10^{14}$\,G.

This paper presents the long-term X-ray monitoring campaign of \aa\ since February 2004. We describe the data reduction in Sect.\,\ref{sec:data}. 
We present the results of the data analysis in Sect.\,\ref{sec:analysis}. Discussion of the results and conclusions follow in Sects.\,\ref{sec:discussion} 
and \,\ref{sec:conclusions}, respectively.

\begin{figure*}
\vspace{-0.2cm}
\begin{center}
\includegraphics[width=1.95\columnwidth]{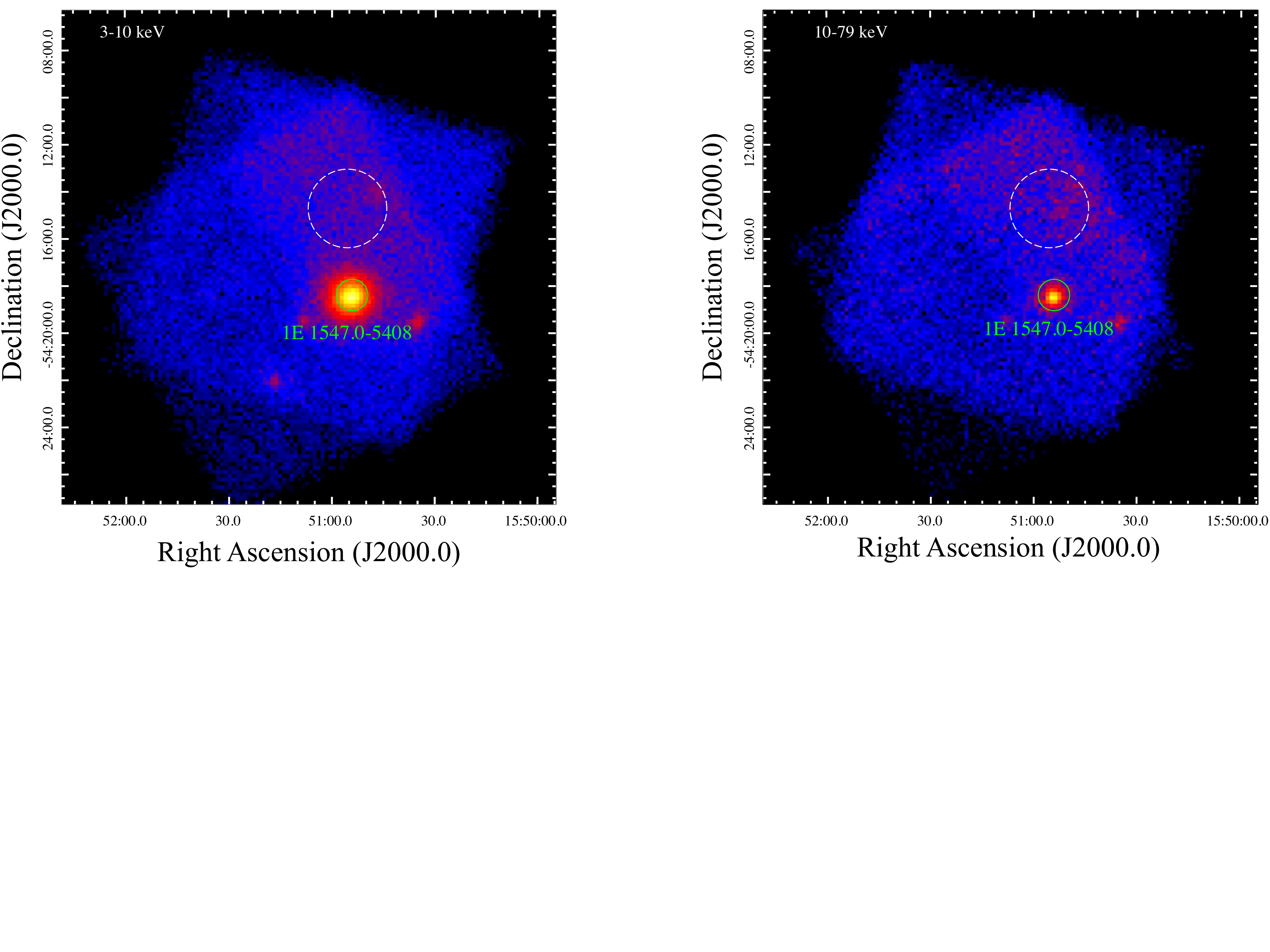}
\end{center}
\vspace{-5.5cm}
\caption{\nustar\ FPMA+FPMB exposure-corrected images combined for the two observations of the field around \aa\ over the 3--10\,keV ({\em left-hand 
panel}) and 10--79\,keV ({\em right-hand panel}) energy bands. North is up, East to the left. The roll angles were 208.3\deg\ and 162.8\deg\ for the first and 
the second observation, respectively, measured from North towards East. Both images were smoothed with a Gaussian filter with a kernel radius of 3 
pixels (one \nustar\ FPM pixel corresponds to about 2.46 arcsec), and the same color code was applied to both cases. The green circle indicates the 
extraction region adopted to collect the source counts, while the dashed white circle marks an example of an extraction region used to estimate the 
background level (see the text for details).}  
\label{fig:fov}
\vskip -0.1truecm
\end{figure*}

\section{Observations and data reduction}
\label{sec:data}

\subsection{Swift observations}

\aa\ was monitored 392 times with the X-ray Telescope (XRT; \citealt{burrows05}) on board \emph{Swift} between 2007 June 22 and 2019 February 
17 for a total exposure of $\sim$\,1.26\,Ms. The single exposures ranged from $\sim$\,0.2 to $\sim$\,15\,ks, with the XRT configured either in photon 
counting mode (PC; time resolution of 2.51\,s) or windowed timing mode (WT; time resolution of 1.77\,ms). The results of the spectral analysis for the 
observations performed until the end of 2011 have already been presented by \cite{cotizelati18}. Here, we focus on the observations performed 
since the beginning of 2012. Only data sets with at least 40 source net counts were retained for the analysis.

Data processing, creation of exposure maps, pile-up corrections, creation of the observation-specific ancillary response files, and extraction of source 
and background spectra were all performed following the standard prescriptions described on the online threads\footnote{\url{http://www.swift.ac.uk/analysis/xrt/index.php}.}. 
A circular region of radius 15 pixels (1 XRT pixel $\sim$\,2.36 arcsec), was adopted to collect the source counts for both PC and WT mode data. For 
WT-mode data, photons with energy $<1$\,keV were discarded owing to known calibration issues at low energies related to this operating mode.

\subsection{NuSTAR observations}

The \nustar\ satellite \citep{harrison13} observed \aa\ twice: on 2016 April 23--24 (observation ID: 30101035002) and on 2019 February 15--17 
(observation ID: 30401008002). The dead-time corrected on-source exposure times were similar in the two pointings, yielding total exposures of 
173.1 and 172.0\,ks for the focal plane module A (FPMA) and B (FPMB), respectively. We processed the event lists and filtered out passages of 
the satellite through the South Atlantic Anomaly using the script \textsc{nupipeline} of the \nustar\ Data Analysis Software (\textsc{nustardas}, version 
1.9.3, distributed along with \textsc{heasoft} v.\,6.25) and the calibration files stored in the most recent version of the calibration database (\textsc{caldb} 
v. 20180419).

Figure\,\ref{fig:fov} shows the exposure-corrected images of the field around \aa\ obtained from the merged FPMA+FPMB event files for the two 
observations over two different energy intervals. Stray light contamination is evident in both datasets. For each of the two observations, we 
accumulated the source photon counts within a circular region of radius 40 arcsec centered on the most accurate position and the background 
counts from different regions (an annulus with inner and outer radii of 80 and 130 arcsec, a closeby circle with the same radius as the source 
extraction region on the same detector chip or a circle of radius 100 arcsec placed on a region affected by high stray light contamination; see the 
dashed white circle in Fig.\,\ref{fig:fov}). We verified that the different choices for the background regions yielded similar results in the following 
analysis. The source average net count rates over the 3--79\,keV energy band were $(0.110\pm0.003)$ counts\,s$^{-1}$ and $(0.114\pm0.002)$ 
counts\,s$^{-1}$ (summing up the two FPMs) during the first and the second observation, respectively; hence, they are consistent with each other 
within the uncertainties.

We applied the \textsc{nuproducts} tool to extract background-subtracted spectra and to generate instrumental response and auxiliary files for each 
FPM and for both observations. All photons outside the 3--79\,keV energy interval were flagged as bad. Given the similarity of the net count rates 
along the two epochs (see above), we decided to combine all the spectra and response files to increase the photon counting statistics and better 
constrain the source spectral shape. The resulting background-subtracted spectrum was then grouped so as to contain at least 20 photons per 
energy bin. For the timing analysis, we referred the photon arrival times of the source event files to the Solar System barycenter using the \textsc{barycorr} 
tool and version 91 of the \nustar\ clock file to correct for drifts of the spacecraft clock\footnote{\url{http://www.srl.caltech.edu/NuSTAR_Public/NuSTAROperationSite/clockfile.php}.}.

\section{Analysis and results}
\label{sec:analysis}

\subsection{Spectral analysis and long-term light curves}

We modeled the \swift\ XRT spectra acquired since the beginning of 2012 within the \textsc{xspec} package (v. 12.10.1; \citealt{arnaud96})  
following the prescriptions reported by \cite{cotizelati18}: we employed an absorbed blackbody (BB) plus power law (PL) model, adopting 
the \textsc{TBabs} model with cross-sections of \cite{verner96} and elemental abundances of \cite{wilms00} to describe the photo-electric 
absorption by the interstellar medium along the line of sight. We fixed the column density to $N_{\rm H} = 4.9 \times 10^{22}$\,cm$^{-2}$ (see 
\citealt{cotizelati18}).

As a first step, we allowed all the parameters to vary among the spectra. The best-fitting values for the BB temperature and radius and the PL 
photon index were found to be compatible with each other over the past $\sim$\,9\,yrs within a 2$\sigma$ confidence level. To better constrain 
the temporal evolution of these parameters, we jointly fit the spectra acquired during the same year, tying up the above-mentioned parameters 
among the data sets. We obtained statistically acceptable results in all cases (the reduced chi-squared $\chi_{\nu}^2$ ranged from 0.89 to 1.05). 
The BB temperature and emitting radius were between $kT_{{\rm BB}}$\,$\sim$\,0.6 -- 0.8\,keV and $R_{{\rm BB}}$\,$\sim$\,0.6 -- 1.3\,km, while 
the PL photon index varied between $\Gamma$\,$\sim$\,2.4 -- 3.3. However, we did not observe any clear trend in the time evolution of these parameters.

\begin{figure}
\begin{center}
\includegraphics[width=1.\columnwidth]{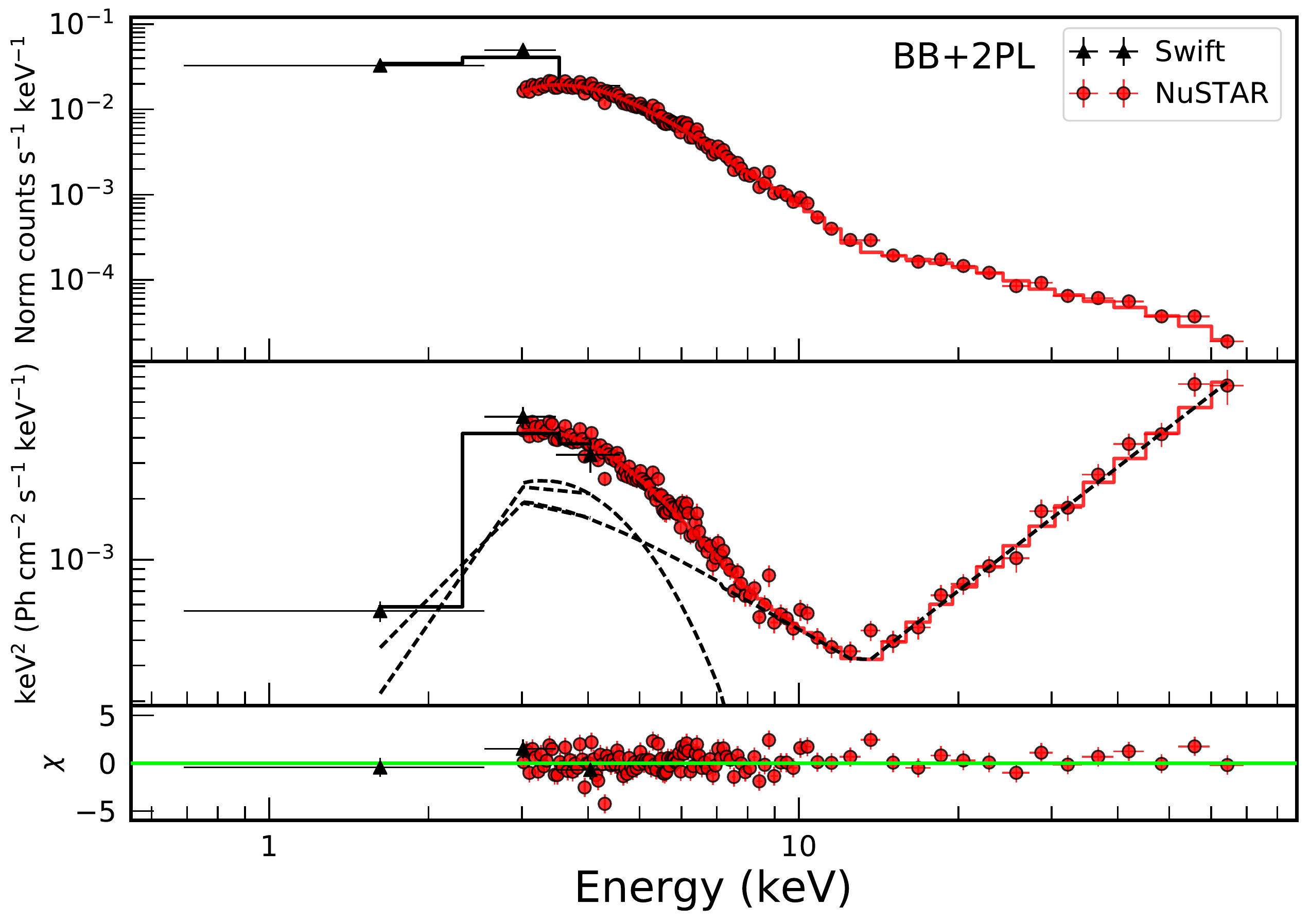}
\end{center}
\vspace{-0.3cm}
\caption{Spectrum of \aa\ extracted over the 0.7--70\,keV energy band from the \swift\ XRT (black triangles) and merged \nustar\ data sets (red 
circles). Data points were re-binned for plotting purpose. The solid line represents the best-fitting BB+2PL model. The $E^2f(E)$ unfolded spectrum 
is shown in the {\em middle panel}. Dashed lines mark the contribution of the single components to the spectral model. Post-fit residuals in units of standard 
deviations are shown in the {\em bottom panel}.}  
\label{fig:spectra}
\vskip -0.1truecm
\end{figure}

The upper panel of Fig.\,\ref{fig:spectra} shows the combined spectra extracted from the \nustar\ and \swift\ XRT data acquired in WT-mode 
close to the epoch of the first \nustar\ observation (obs ID: 00030956189; exposure time of $\sim1.7$\,ks). A change in the spectral slope can 
be clearly seen at an energy of $\approx$\,15\,keV, calling for an additional spectral component at higher energies. Hence, we fitted different 
multi-component models to the \nustar\ plus \swift\ spectrum: a double-BB plus a PL model (2BB+PL), a BB plus a broken PL model (BB+2PL), 
and the resonant Compton scattering model \textsc{ntz} (see \citealt{nobili08}) plus a PL component (\textsc{ntz}+PL). The hydrogen column 
density was fixed to the values we obtained for the outbursts in 2008 and 2009 under the same assumptions for the absorption model 
\citep{cotizelati18}, that is, $N_{\rm H}=4.9\times10^{22}$\,cm$^{-2}$ for the BB+2PL model and $N_{\rm H}=4.6\times10^{22}$\,cm$^{-2}$ 
for the 2BB and \textsc{ntz}+PL models. A renormalization constant was also included to account for intercalibration uncertainties, and was found 
to be consistent within the 1$\sigma$ uncertainties across the different instruments in all cases.

\begin{table*}
\begin{center}
\caption{Results of the joint fits of the \swift\ spectrum and the combined spectrum extracted from the two \nustar\ observations of \aa.}
\label{tab:spectra}
\resizebox{2.05\columnwidth}{!}{
\begin{tabular}{@{}lccccccccc}
\hline
Model				& $kT_{{\rm BB_1}}$ 	& $R_{{\rm BB_1}}$			& $kT_{{\rm BB_2}}$		& $R_{{\rm BB_2}}$		& $\Gamma_{\rm hard}$	& $F_{{\rm X,abs}}$		& $L_X$ 			& $L_{{\rm BB_1}}$ / $L_{{\rm BB_2}}$ / $L_{{\rm PL,H}}$			& $\chi_{\nu}^2$ (d.o.f.)	\\	
					& (keV)				& (km)					& (keV)				& (km)				&					& ($10^{-11}$ \flux)		& ($10^{34}$ \lum)	& ($10^{34}$ \lum)							    &	\\ \hline
\textsc{2BB+PL} 		& $0.58\pm0.03$		& $1.7\pm0.4$				& $1.13\pm0.08$		& $0.2\pm0.1$ 			& $0.00\pm0.09$	 	& $1.53\pm0.07$	    	& $6.6\pm0.1$ 	    	& $(3.97\pm0.07)$ / $(0.8\pm0.1)$ / $(1.8\pm0.1)$	& 0.97 (301)	 \\  \hline \hline
    					& $kT_{{\rm BB}}$ 			& $R_{{\rm BB}}$				& $\Gamma_{\rm soft}$ & $E_{{\rm break}}$		& $\Gamma_{\rm hard}$	& $F_{{\rm X,abs}}$		& $L_X$			& $L_{{\rm BB}}$ / $L_{{\rm PL,S}}$ / $L_{{\rm PL,H}}$	  						 				    & $\chi_{\nu}^2$ (d.o.f.)	\\	
					& (keV)				& (km)					&					& (keV)				&					& ($10^{-11}$ \flux) 		& ($10^{34}$ \lum)	& ($10^{34}$ \lum)							 			    &	\\ \hline
\textsc{BB+2PL}		& $0.66\pm0.01$		& $0.9\pm0.3$				& $3.8\pm0.2$			& $12.8\pm0.4$		& $0.0\pm0.1$			& $1.60\pm0.06$	    	& $7.0\pm0.2$ 		& $(2.6\pm0.2)$ / $(2.7\pm0.3)$ / $(1.7\pm0.2)$									    	& 0.97 (301)	 \\  \hline \hline						  	
					& $kT$ 				& $\beta$			    		& $\Delta\psi$ 			& norm				& $\Gamma_{\rm hard}$	& $F_{{\rm X,abs}}$		& $L_X$ 			& $L_{\textsc{ntz}}$ / $L_{{\rm PL,H}}$				& $\chi_{\nu}^2$ (d.o.f.)	\\	
	     				& (keV)				&						& (rad)				& (10$^{-1}$)			&					& ($10^{-11}$ \flux) 		& ($10^{34}$ \lum)	& ($10^{34}$ \lum)							    &	\\ \hline
\textsc{ntz+PL}			& $0.61\pm0.01$		& $0.20\pm0.01$			& $0.71\pm0.02$		& $1.2\pm0.1$			& $-0.20\pm0.08$		& $1.56\pm0.05$	    	& $6.8\pm0.1$		& $(4.78\pm0.05)$ / $(1.9\pm0.1)$				& 0.99 (301)	 \\  \hline \vspace{0.1cm} 
\end{tabular}
}
\end{center}
\small
{\bf Notes.} The absorption column density was fixed to $N_{\rm H}=4.6\times10^{22}$\,cm$^{-2}$ for the 2BB+PL and \textsc{ntz}+PL models, and 
to $N_{\rm H}=4.9\times10^{22}$\,cm$^{-2}$ for the BB+2PL model. In the \textsc{ntz} model, $\beta$ denotes the bulk motion velocity of the charged 
particles in the magnetosphere and $\Delta\psi$ the twist angle.
BB radii and luminosities are evaluated for an observer at infinity and for a source distance of 4.5\,kpc. $F_{{\rm X,abs}}$ and $L_X$ are evaluated 
within the 0.3--70\,keV energy range; $L_{{\rm BB_1}}$, $L_{{\rm BB_2}}$, $L_{{\rm BB}}$, $L_{{\rm PL,S}}$ and $L_{\textsc{ntz}}$ in the 0.3--10\,keV 
interval; $L_{{\rm PL,H}}$ within the 10--70\,keV band. In the BB+2PL model, the soft PL component tends to give an unphysical overestimate for the 
luminosity below $\approx$\,1\,keV. In order to obtain a more reliable estimate for $L_{{\rm PL,S}}$, the soft PL component is actually a broken PL 
where we fix the photon index of the low-energy PL component to $-$3, and allow the break energy to vary. The best-fitting value for the break energy is 
around 1\,keV, so that the contribution to $L_{{\rm PL,S}}$ is negligible below 1\,keV.
\end{table*}

\begin{figure*}
\hspace{-0.5cm}
\begin{center}
\includegraphics[scale=0.35]{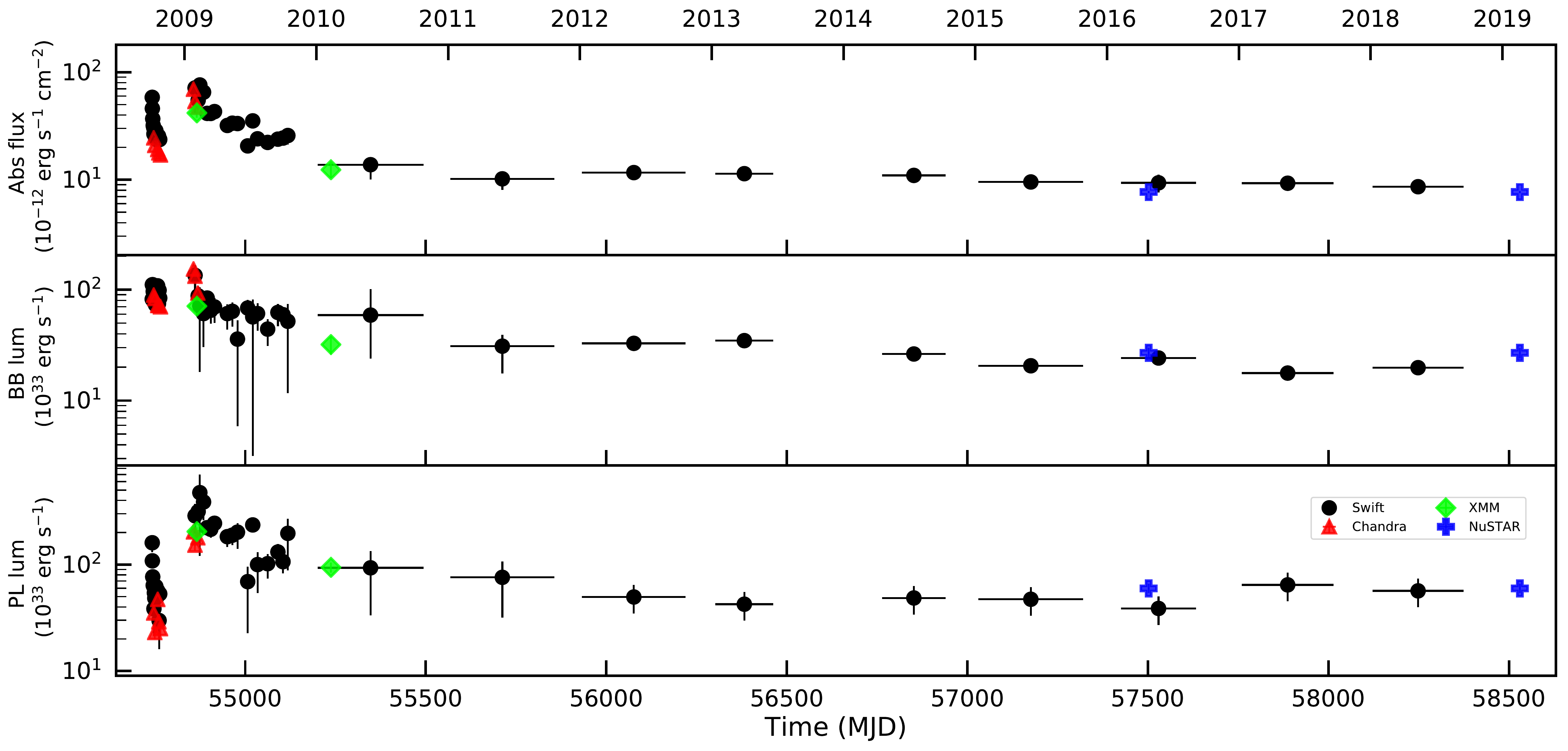}
\includegraphics[scale=0.27]{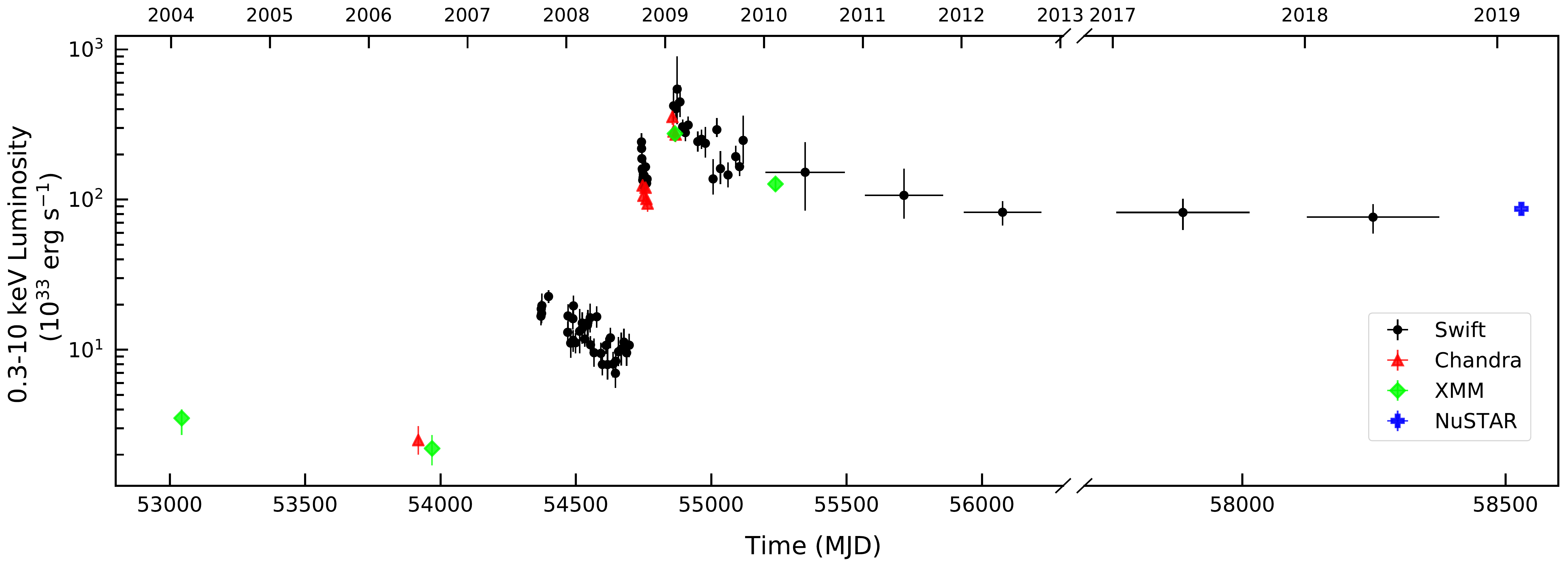}
\end{center}
\vspace{-0.3cm}
\caption{{\em Top}: long-term evolution of the observed flux and of the luminosities of the BB and soft PL spectral components of \aa\ between 2008 
October 3 and 2019 February 15. {\em Bottom}: long-term evolution of the soft X-ray luminosity of \aa\ between 2004 February 8 and 2019 February 
15. The values derived from \cxo\ and \xmm\ observations are taken from \cite{cotizelati18}. All quantities are evaluated over the 0.3--10\,keV 
energy band. Luminosities have been estimated following the same procedure outlined in Table\,\ref{tab:spectra}. The luminosities in the time interval 
between 2013 and 2017 are consistent with those measured after 2017, and are not shown for plotting purpose.}  
\label{fig:lcurve}
\vskip -0.1truecm
\end{figure*}

All three models provided satisfactory results ($\chi_{\nu}^2=0.97$, 0.97 and 0.99 for 301 d.o.f., for the 2BB+PL, BB+2PL and \textsc{ntz}+PL models, 
respectively). The best-fitting values for the spectral parameters, the fluxes and the luminosities are listed in Table\,\ref{tab:spectra}. Figure\,\ref{fig:spectra} 
shows the broad-band spectrum fitted with the BB+2PL model, the unfolded spectrum highlighting the contribution of the different components, and 
the post-fit residuals. We derived fully compatible results when modeling the \nustar\ spectrum together with the \swift\ XRT spectra acquired in PC 
mode in observations nearly simultaneous to the second \nustar\ observation (IDs: 00088680002, 00088680003 for a total exposure time of 3.3\,ks). 
The observed fluxes in the 1--10\,keV and 15--60\,keV energy bands are $F_{1-10}$\,$=$\,$7.5^{+0.2}_{-1.0}$\,$\times$\,$10^{-12}$ \flux\ and 
$F_{15-60}$\,$=$\,$4.6^{+0.1}_{-0.6}$\,$\times$\,$10^{-12}$ \flux, respectively (here and in the following, all uncertainties are quoted at a confidence 
level (c.l.) of 1$\sigma$). The flux hardness ratio, $\eta$\,$=$\,$F_{15-60}/F_{1-10}$\,$\approx$\,0.6, is smaller than that obtained during past \su\ 
observations in 2009 and 2010, $\eta$\,$=$\,$1.2-1.8$ (see \citealt{enoto17}). However, it still follows the correlation between the spectral hardness 
and the dipole magnetic field reported by \cite{enoto17}. To derive a constraint on the steepening of the hard PL component at high energies, we 
restricted our spectral analysis to energies above 13\,keV, and fitted a PL model with high-energy exponential cutoff (\textsc{cutoffpl} in \textsc{xspec}) 
to the data. We inferred a lower limit on the high-energy roll-over of  $E_{{\rm cut}} > 260$\,keV (3$\sigma$ c.l.).

Figure\,\ref{fig:lcurve} shows the long-term evolution of the observed flux and of the luminosities for the single spectral components as well as the 
total one in the soft X-ray band. \aa\ did not return to the level observed in 2006 following the three outbursts in 2007, 2008 and 2009. In particular, 
the X-ray flux over the past $\sim$\,9\,yr has settled on a value of $\approx$\,$10^{-11}$ \flux. This is about an order of magnitude below the value 
at the peak of the 2009 outburst, a factor of $\sim$\,5 larger than those measured in 1980 and 1998, and a factor of $\sim$\,30 above that detected 
in July 2006.

\begin{figure*}
\begin{center}
\includegraphics[width=1.9\columnwidth]{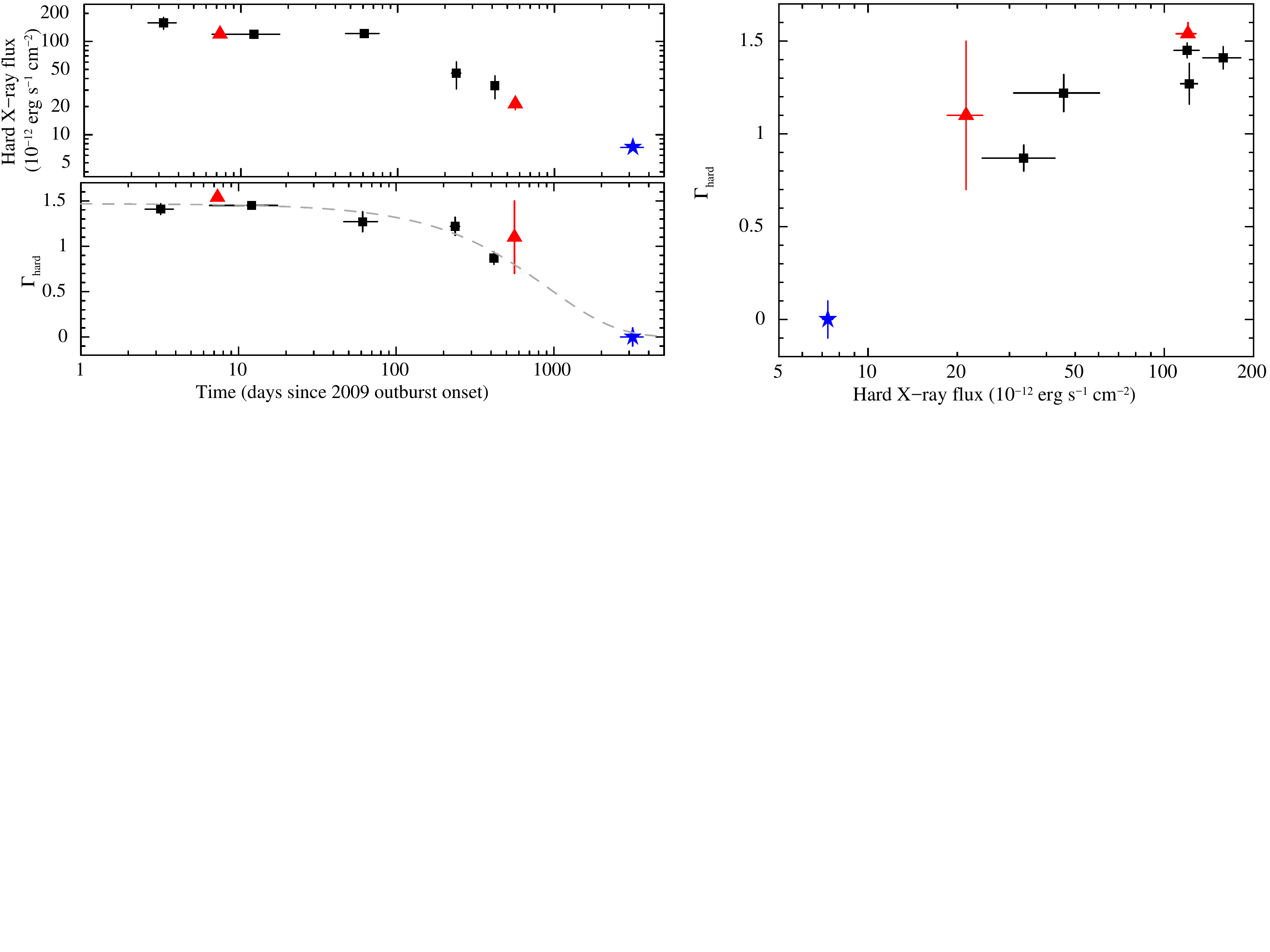}
\end{center}
\vspace{-7.6cm}
\caption{{\em Left}: time evolution of the flux in the 10--70\,keV energy band ({\em top}) and of the PL photon index in the hard X-rays ({\em bottom}) of \aa\ since the 
onset of the last outburst on 2009 January 22 at 00:53 UTC. Black squares refer to \integral\ data, red triangles refer to \su\ data, the blue star 
refers to the merged \nustar\ observations presented in this study. All fluxes for archival observations were referred to the 10--70\,keV band using \textsc{pimms} 
(\texttt{https://heasarc.gsfc.nasa.gov/cgi-bin/Tools/w3pimms/w3pimms.pl}), assuming the spectral shapes reported in Table\,3 by \cite{enoto10}, Table\,8 by 
\cite{kuiper12} and Table\,3 by \cite{iwahashi13}. The dashed grey line in the bottom panel denotes the best-fitting exponential function for the time evolution 
of the photon index (see the text for details). {\em Right}: evolution of the PL photon index for the hard X-ray component as a function of the hard X-ray flux. 
Marks and colors are the same as in the {\em left panel}.}  
\label{fig:pl_evolution}
\vskip -0.1truecm
\end{figure*}

The left-hand panel of Fig.\,\ref{fig:pl_evolution} shows the time evolution of the flux (top) and spectral shape for the PL component (bottom) in the 
hard X-ray band since the onset of the 2009 outburst (when it was first detected), extracted using the values published by \cite{enoto10}, \cite{kuiper12} 
and \cite{iwahashi13} for the archival observations. The hard X-ray flux measured in the recent \nustar\ observations, $\sim$\,7\,$\times$\,$10^{-12}$ 
\flux\ over the 10--70\,keV energy band, is a factor of $\approx$\,20 smaller than that measured at the peak over the same band. The PL gradually 
hardened in time, with the photon index decreasing from $\Gamma_{\rm hard}$\,$\sim$\,1.4 about 3 days after the 2009 outburst onset \citep{kuiper12}, 
to $\Gamma_{\rm hard}$\,$\sim$\,0 about ten years later (see also Table\,\ref{tab:spectra}). We estimated the rate of the PL hardening by fitting an 
exponential function of the form $\Gamma_{\rm hard}(t)=\Gamma_0~ \rm{exp}(-$\emph{t}$/\tau_h)$ to the time evolution of the photon index ($t$ 
represents the time since the outburst onset and $\tau_h$ the $e$-folding time). We obtained a good description of the data ($\chi^2_\nu = 1.18$ for 
6 d.o.f.; see the dashed light grey line in Fig.\,\ref{fig:pl_evolution}), and derived best-fitting parameters $\Gamma_0=1.45\pm0.03$ and 
$\tau_h=879^{+162}_{-124}$\,d. All in all, the decrease of the hard X-ray flux in time appears to be accompanied by a hardening of the spectrum 
at high energies (see the right-hand panel of Fig.\,\ref{fig:pl_evolution}).

\begin{figure}
\vspace{-0.1cm}
\begin{center}
\includegraphics[scale=0.4]{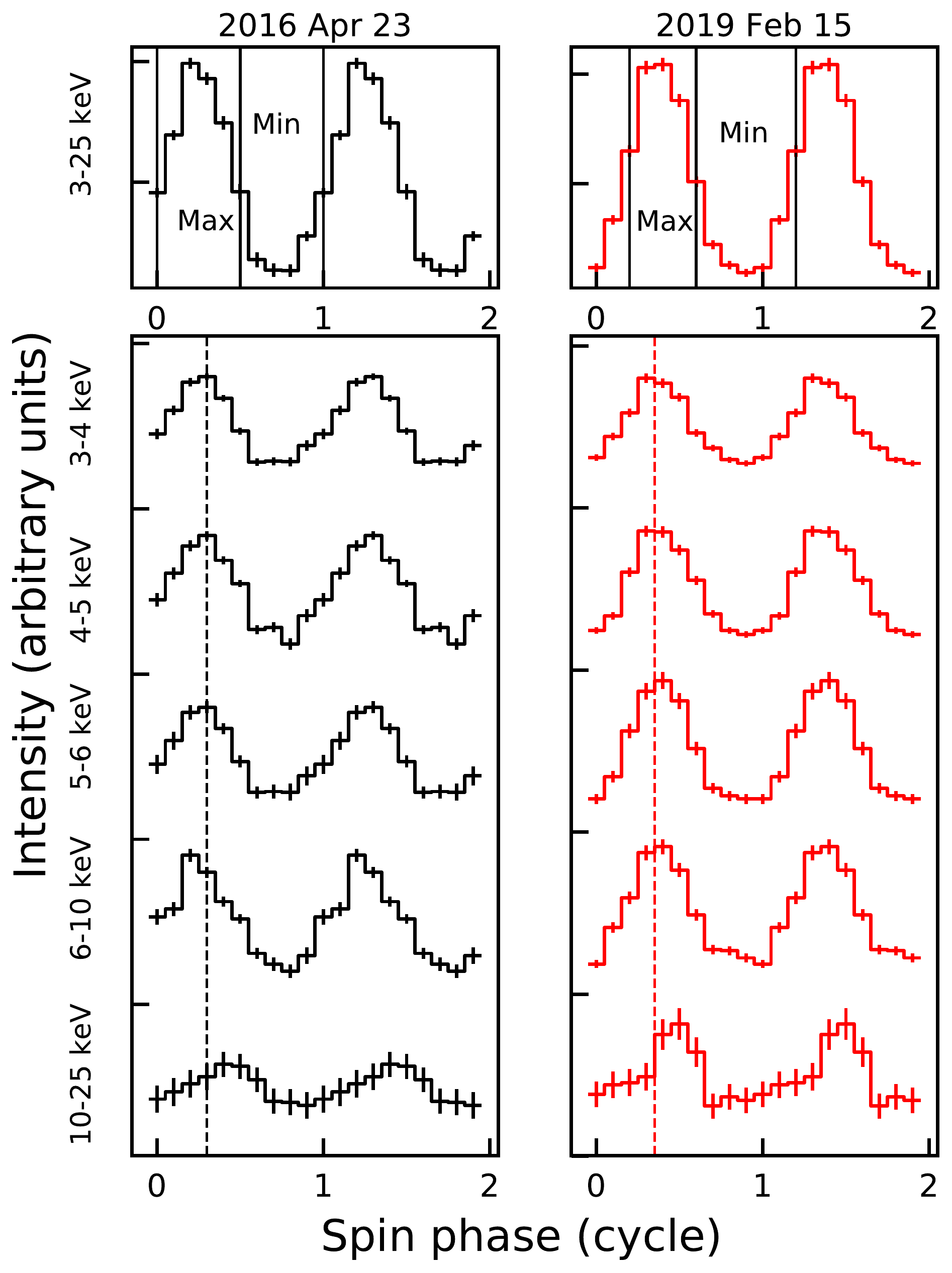}
\includegraphics[scale=0.37]{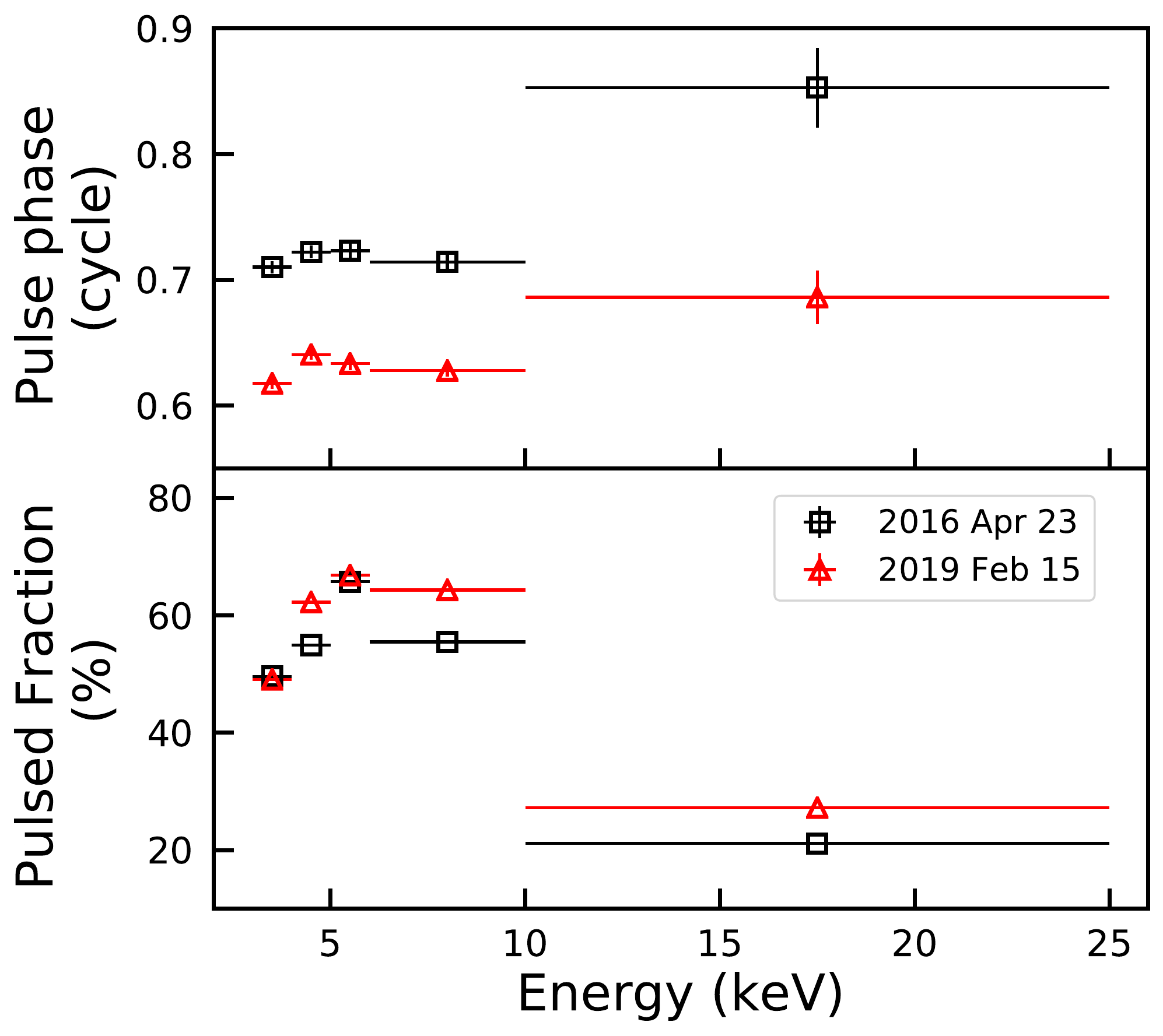}
\end{center}
\vspace{-0.3cm}
\caption{
\emph{Top}: background-subtracted pulse profiles of \aa\ extracted over the 3--25\,keV energy range from the combined \nustar\ FPMA and 
FPMB data sets, and separately for both observations. The vertical black solid lines mark the phase intervals adopted to extract the phase-resolved 
spectra. \emph{Middle}: pulse profiles extracted in selected energy ranges. The profiles have been shifted along the vertical axis for plotting purpose. 
The vertical dashed lines mark the phase of the pulse peak in the soft X-ray energy bands (up to 10\,keV). All pulse profiles in the {\em top} and 
{\em middle panels} are sampled in 10 phase bins, and two phase cycles are shown. For display purposes, the 2016 pulse profiles have been phase-aligned 
to the 2019 ones. \emph{Bottom}: pulse phase ({\em top}) and pulsed fraction ({\em bottom}) as a function of energy at both epochs.}
\label{fig:timing}
\vskip -0.1truecm
\end{figure}

\subsection{Pulse profiles and phase-resolved spectral analysis for the NuSTAR observations}
\label{sec:timing}

We computed the $Z^2_2$ test \citep{buccheri83} on the source event lists from the combined FPMA and FPMB data sets separately for the 
two observations. We detected a prominent peak at a period of $P$\,$\sim$\,$2.08671$\,s in the first observation and $P$\,$\sim$\,$2.08843$\,s in 
the second observation. We honed these values by dividing the time series of both observations in 6 time segments of approximately equal length, 
and applying a phase-fitting technique. Fitting a first order polynomial function to the phase evolution yielded $P$\,$=$\,$2.0867100(1)$\,s at 
$T_{0,1st}$ = 57501.026\,MJD and $P$\,$=$\,$2.0884305(3)$\,s at $T_{0,2nd}$ = 58529.107\,MJD. This implies a net spin-down rate of 
$\sim$\,1.9\,$\times$ 10$^{-11}$\,s\,s$^{-1}$ between April 2016 and February 2019.

The top and middle panels of Fig.\,\ref{fig:timing} show the background-subtracted light curves extracted from the combined FPMA and FPMB 
data sets, folded on our best value for $P$ measured in each observation. We detected pulsed emission up to $\approx$\,25\,keV. The profile shape
above 10\,keV is apparently different from that observed at lower energies. Moreover, the peak of the pulse profile at higher energies appears to 
lag the peak at lower energies. 

To evaluate the shift as well as the pulsed fraction values as a function of photon energy, we modeled the 
pulse profiles with a constant plus two sinusoidal functions, fixing the sinusoidal periods to those of the fundamental and second harmonic 
components. The inclusion of higher harmonic components in the fit was not statistically needed ($F$--test probability of 0.48 for the improvement in the fit). 
We evaluated the phase lags as the difference in the pulse phase of 
the fundamental component between the pulse profiles extracted over the 10--25\,keV and 3--4\,keV energy intervals (see the bottom panel of Fig.\,\ref{fig:timing}).
We determined a phase lag of $\Delta\phi_1 = 0.15\pm0.03$ cycles in April 2016 and $\Delta\phi_2 = 0.07\pm0.02$ cycles in February 2019.  
We obtained compatible values when considering the 4--5\,keV, 5--6\,keV and 6--10\,keV energy intervals as a reference for the soft X-ray band.
We computed the pulsed fraction in the different energy intervals by dividing the value of the semi-amplitude for the fundamental component of the 
pulse profile by the average count rate. In both observations, the pulsed fraction shows an increasing trend up to 6\,keV, from $\sim$\,49\% in the 
3--4\,keV interval up to $\sim$\,65\% in the 5--6\,keV range. Then, it decreases at higher energies, down to $\sim$\,20-25\% over the 10--25\,keV 
energy range (see the bottom panel of Fig.\,\ref{fig:timing}). Pulsations are no more significantly detected at higher energies. We estimate 4$\sigma$ 
upper limits on the pulsed fraction of 23\,\% and 14\,\% for the first and the second observation, respectively, over the 25--70\,keV energy range.

The pulsed fraction at low energies is larger than that measured over the 1--6\,keV energy range in 2006 ($\approx$\,15\,\%; \citealt{halpern08}).
We reanalyzed the data taken in a 12-ks observation with \cxo\ ACIS-S (ID: 10186) on 2009 February 2, $\sim$\,10\,d after the outburst onset. We 
detected the spin signal at $P=2.07211(1)$\,s, and found pulsed fractions of $\sim$\,11\,\% in the 3--4 and 4--5\,keV ranges, $\sim$\,5\,\% in the 
5--6\,keV interval and $\sim$\,8\,\% in the 6--10\,keV range. Hence, the pulsed fraction increased considerably along the outburst decay in the 
soft X-ray energy bands. On the other hand, the detection of pulsed emission in the hard X-ray band appears quite surprising considering the 
decreasing trend of the pulsed fraction along the first year of the outburst, down to undetectable levels about 11 months after the outburst 
onset (\citealt{kuiper12}; see e.g. their Fig.\,9), when the flux was a factor of $\sim$\,4.5 larger than the value measured 
in the recent \nustar\ observations (see the top panel of Fig.\,\ref{fig:pl_evolution}).

\begin{figure}
\begin{center}
\includegraphics[width=.98\columnwidth]{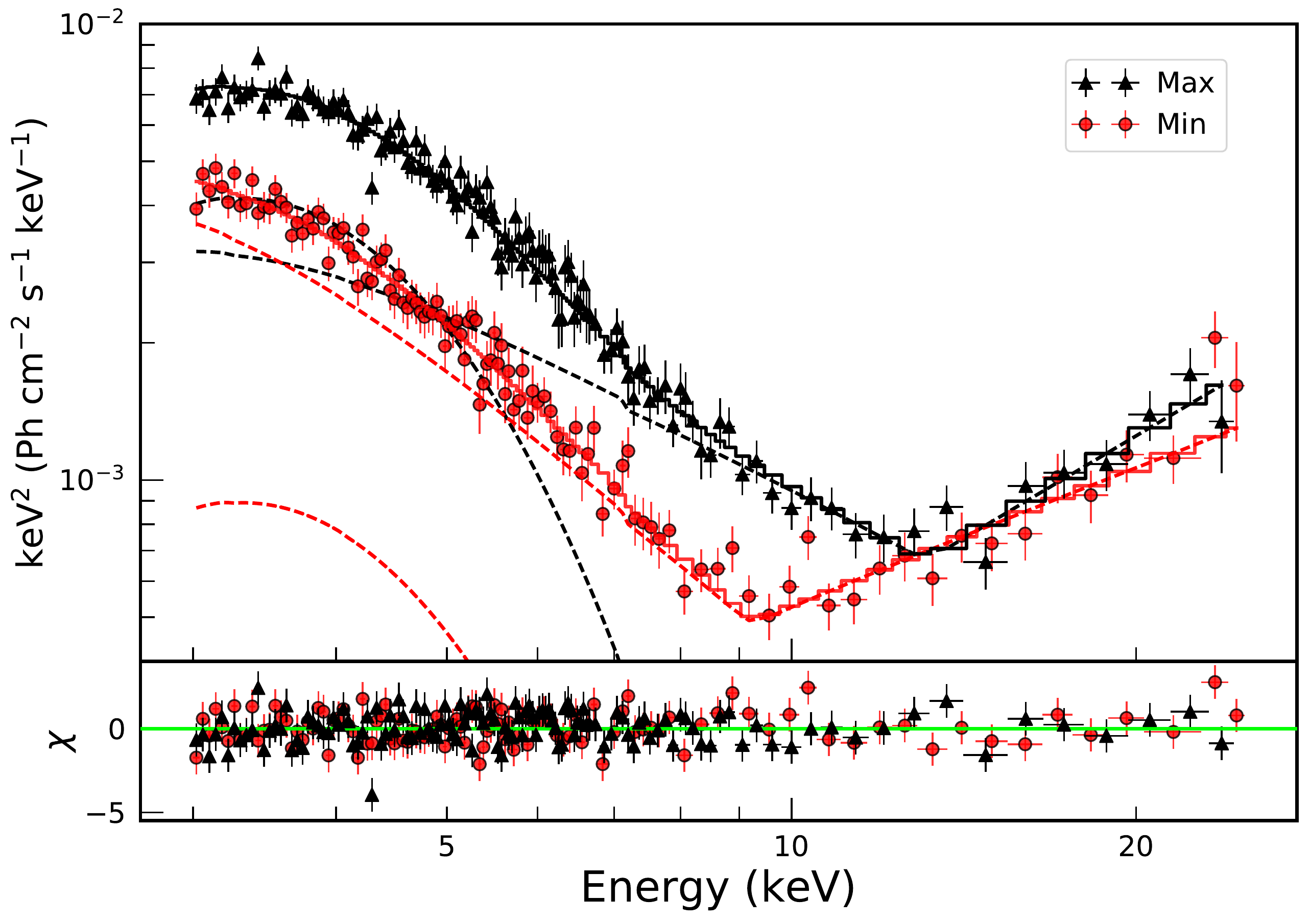}
\end{center}
\vspace{-0.3cm}
\caption{{\it Top}: $E^2f(E)$ unfolded spectra of \aa\ extracted from the merged \nustar\ data sets at the maximum (in black) 
and at the minimum (in red) of the pulse profile. Data points were re-binned for plotting purpose. The solid lines represent the best-fitting BB+2PL model. 
Dashed lines mark the contribution of the single components to the spectral model. {\it Bottom}: post-fit residuals in units of standard deviations.}  
\label{fig:spectra_pps}
\vskip -0.1truecm
\end{figure}

We performed a pulse phase-resolved spectral analysis for both \nustar\ observations using data from both FPMs. For each observation, we 
extracted the spectra within the phase intervals corresponding to the maximum and the minimum of the pulse profiles (these intervals were selected 
as indicated in the top panel of Fig.\,\ref{fig:timing}). We then combined the spectra for the maximum and the minimum from the two observations to 
increase the photon counting statistics\footnote{Very similar parameter values were derived from the spectral analysis of the datasets separately at 
the two different epochs.}, and fitted them together using an absorbed BB+2PL model (the column density was fixed at the phase-averaged value). 
The BB temperatures were found to be consistent with each other and with the phase-averaged value, and were thus fixed to that value in the 
spectral fits. We obtained an acceptable result, with $\chi_{\nu}^2 = 0.95$ (284 d.o.f.). The spectral shape changes significantly as a function of the 
rotational phase (see Fig.\,\ref{fig:spectra_pps}). We derived the following parameters: 
$\Gamma_{\rm soft, max} = 3.4\pm0.1$, $\Gamma_{\rm hard, max} = 0.5\pm0.3$, $E_{{\rm break, max}} = (13.1\pm0.6)$\,keV at the maximum;  
$\Gamma_{\rm soft, min} = 4.20\pm0.09$, $\Gamma_{\rm hard, min} = 1.0\pm0.1$, $E_{{\rm break, min}} = (9.2\pm0.2)$\,keV at the minimum.

\section{Discussion}
\label{sec:discussion}

\subsection{A very slow luminosity decay?}

The long-term light curve of \aa\ in the soft X-ray band since the onset of its last outburst in 2009 reveals an initial decay by one order of magnitude
in flux over the first year (see Fig.\,\ref{fig:lcurve}). Such a decay pattern is typically observed for magnetar outbursts \citep{cotizelati18}. However, 
in the past 9 years, the source has always maintained a flux of $\approx$\,$10^{-11}$ \flux\, which is a factor of $\sim$\,30 above the value measured 
in 2006, and a factor of $\sim$\,5 larger than the values observed in 1980 and 1998. Such a prolonged, possibly persistent, high X-ray flux is a property 
that, so far, has not been seen so clearly in other magnetars. The long-lasting high temperature ($kT_{{\rm BB}}$\,$\gtrsim$\,0.6\,keV) is an additional 
feature that so far has been observed over a few years only in other few cases, e.g. during the recent outbursts of the magnetars SGR\,J1745$-$2900 
\citep{cotizelati15,cotizelati17} and \wes\ \citep{borghese19}.

On the other hand, the conspicuous softening of the PL component in the soft X-rays (from $\Gamma$\,$\sim$\,1.2 at the outburst peak to 
$\Gamma$\,$\sim$\,3.8 about a decade later), the reduction of the BB emitting radius (from $R_{{\rm BB}}$\,$\sim$\,3\,km down to $R_{{\rm BB}}$\,$\sim$\,1\,km) 
and the weakening of the hard X-ray tail are all properties commonly seen in magnetars along the outburst decay (e.g., \citealt{kaspi17,esposito18}). 
The current broad-band X-ray properties of \aa\ are also very similar to those observed in other magnetars. Changes in the profile structure with energy 
are observed in \rxs\ \citep{gotz07,denhartog08}, \kes\ \citep{an15}, \ea\ \citep{vogel14}, \uu\ \citep{tendulkar15}, \ee\ \citep{yang16}, \xte\ \citep{gotthelf19}, 
\sgrb\ \citep{tamba19}. The pulse profiles of \rxs, \ea\ and \xte\ also display a phase shift with energy. All these magnetars also show some degree of 
variability of the pulsed fraction with energy, as well as of the hard X-ray spectral shape along the rotational phase cycle.

In the following, we focus on the unusual long-term evolution of the X-ray flux of \aa\ over the last decade. In a recent systematic study of magnetar 
outbursts \citep{cotizelati18}, we characterized the post-outburst luminosity trend of a large sample of magnetars using a simple phenomenological 
model. In that work, the luminosity decay was fit using an exponential function with two free parameters: 
\begin{equation}
\label{eq:dec}
L(t)= (L_{{\rm max}} - L_0)~e^{-t/\tau} + L_0~,
\end{equation}
where $L_{\rm max}$ is the luminosity at the outburst peak, $L_0$ is the value to which the luminosity tends to decay, $t$ is the time elapsed since the epoch 
of the outburst peak, and $\tau$ is the $e$-folding time. It was assumed that $L_0$ corresponds to the pre-outburst luminosity (i.e., that each source 
returns back to the pre-outburst level, and that $L_0$ is always the same for a given source). The simple model above was used to predict the decay timescale, 
that is how long the luminosity takes to return to a value compatible with that before the outburst (the luminosity measured in 2006 for the case of \aa), as well 
as the total extra energy released during the outburst $E_{\rm out} = \tau (L_{{\rm max}}-L_0)$.

In a few cases, the functional form provided by Eq. (\ref{eq:dec}) was too simple and a second exponential function was needed to fit adequately the data:

\begin{equation}
\label{eq:dec2}
L(t)= (L_{{\rm max}} - L_0)~[e^{-t/\tau_1}+e^{-t/\tau_2}] + L_0~.
\end{equation}

This is also the case for the 2009 outburst of \aa, if we apply the same procedure: once  $L_0$ is fixed to $2.2\times10^{33}$ \lum\ (i.e., the value measured 
in 2006; see Fig.\,\ref{fig:lcurve}), the best-fitting decay model is represented by a double-exponential function with $\tau_1=164\pm43$\,d and $\tau_2=8040^{+5735}_{-3780}$\,d 
(yielding $\chi^{2}_{\nu}=1.22$ for 287 d.o.f.).

\begin{figure}
\begin{center}
\includegraphics[width=1.0\columnwidth]{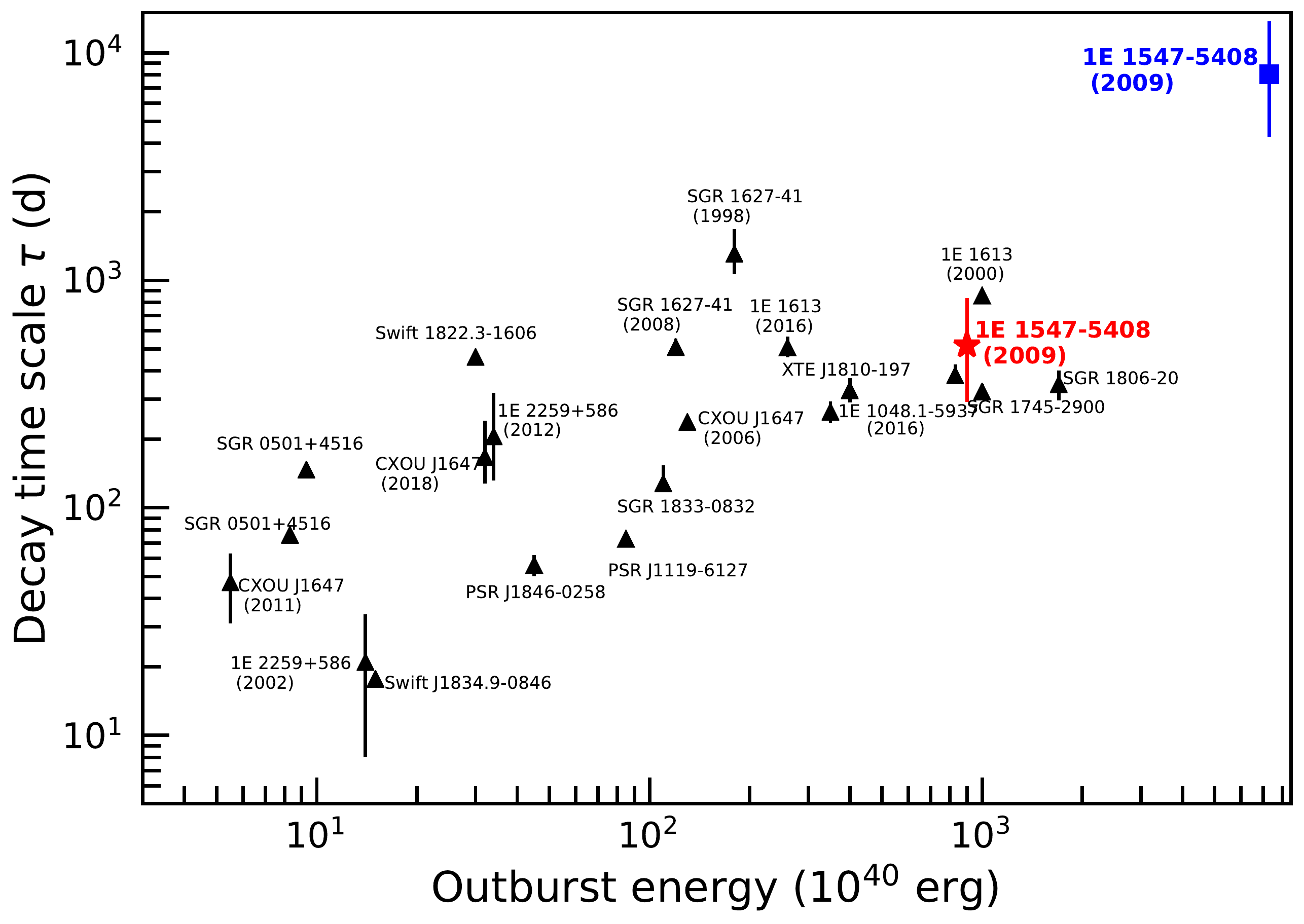}
\end{center}
\vspace{-0.3cm}
\caption{Decay timescale (in terms of the $e$-folding time $\tau$) as a function of the total energy released for all magnetar outbursts detected so far. For 
the cases where more exponential functions were adopted to model the light curve, the value of $\tau$ corresponding to the exponential function modeling 
the late-time evolution of the outburst was considered. The blue square refers to the case where \aa\ is assumed to return to the 2006 level. The red star 
refers instead to the case where \aa\ is assumed to have settled on a new persistent state over the last years. In a few cases, the size of the marker is larger 
than the error bars. Adapted from \cite{borghese19}.}  
\label{fig:corr}
\vskip -0.1truecm
\end{figure}

Figure\,\ref{fig:corr} shows the energy emitted from magnetars during powerful outbursts as a function of the outburst decay timescale. 
Based on the extrapolation of the model above (i.e., assuming no changes in the decay pattern), we estimated that \aa\ would return to a luminosity 
similar to that measured in 2006 (within $\lesssim$\,10\%) in $\approx$\,50\,yr, releasing an energy of $E_{\rm out}$\,$\sim$\,$7.3\times10^{43}$\,erg.
These characteristics would make the 2009 event from \aa\ by far the longest magnetar outburst ever observed (see the blue square in 
Fig.\,\ref{fig:corr}). For a comparison, the second longest outburst hitherto detected according to our models is the 1998 episode from \sgrd, for which the 
long-term light curve can be described by a double-exponential function with $\tau_1$\,$\sim$\,234\,d and $\tau_2$\,$\sim$\,1310\,d. The value of the 
outburst energy for \aa\ would be also exceptionally large, even larger than that estimated for \sgra\ after the giant flare in December 2004, that is 
$\sim$\,$2\times10^{43}$\,erg.\footnote{The value estimated for \sgra\ represents, however, only a lower limit, because the source flux was already 
increasing by a factor of $\sim$\,2 during the first half of 2004 with respect to the level observed previously \citep{mereghetti05}.}

The results obtained with the simple model above rely on the assumption that the value for $L_0$ is equal to that observed before the outburst, meaning that 
magnetars are characterized by a quiescent state which remains the same over decades. However, there is virtually no observation for any magnetar 
which guarantees that their persistent luminosity remains actually constant over these time scales. If we relax the assumption of a fixed $L_0$ in the 
light curve modeling of \aa, we obtain $\chi^{2}_{\nu}=1.14$ for 286 d.o.f.. The $F$-test yields a null hypothesis probability for the improvement of the 
fit of $\simeq6\times10^{-6}$, implying that a model where $L_0$ is allowed to vary provides a more accurate description of the luminosity evolution. 
The best-fitting parameters are $\tau_1=54^{+38}_{-25}$\,d, $\tau_2=518^{+316}_{-226}$\,d, $L_0=(6\pm1)\times10^{34}$\,\lum\ and $E_{\rm out}$\,$\sim$\,$9\times10^{42}$\,erg. 
Hence, in this case, both the decay timescale and the energetics greatly reduce compared to the fit outlined above. These would be more in line with the 
values estimated for other magnetar outbursts (see the red star in Fig.\,\ref{fig:corr}).

The assumption of a fixed $L_0$ is tightly connected to the physical mechanisms that trigger and sustain the outburst activity of magnetars, as 
described in the following sections.

\subsection{Crustal cooling or coronal activity?}

The emission observed during magnetar outbursts has been commonly interpreted as the result of the progressive cooling of a thermally-emitting hot spot 
on the star surface. The hot spot can be formed, arguably, owing to two different, but not mutually excluding, mechanisms. The first one is the sudden 
release of heat in the crust due to the long-term building-up of magnetic stresses in the crust, causing either a starquake (e.g. \citealt{perna11,pons11,pons12,thompson17})
or a thermoplastic wave \citep{beloborodov14,li16}. The second mechanism is the Ohmic dissipation and or particle bombardment upon 
the star surface due to the currents circulating in the magnetar corona (e.g. \citealt{beloborodov09,beloborodov13b,beloborodov13a,beloborodov16,chen17}).

On the one hand, in the crustal cooling scenario, the persistent luminosity $L_0$ is expected to arise from the internal heat diffusion, which yields changes 
over timescales $\gtrsim$\,10$^3$--10$^4$\,yr (assuming a kyr-aged magnetar; see e.g. \citealt{vigano13,potekhin15}) that are longer than any 
observable time span. Hence, this scenario predicts that each source should approximately recover the pre-outburst temperature and thermal luminosity. 
According to this scenario, the size of the emitting region should increase in time due to heat spreading across the crustal layers and the envelope. In 
general, crustal cooling from a single heat release cannot sustain long-lived (several years) surface temperatures of $kT$\,$\gtrsim$\,0.5\,keV. Hence, to 
explain the behaviour of \aa\ in terms of crustal cooling alone, a repeated or continuous heating would be required. Such a heating should be shallow, 
otherwise an enormous input energy would be needed. However, crustal cooling alone cannot account for non-thermal hard X-ray emission, such as that 
observed in \aa.

On the other hand, in the coronal currents scenario, the currents are responsible for both thermal X-ray emission, via their slow dissipation in localized 
regions close to the surface of the star, which is highly resistive (e.g. \citealt{gonzalez19}), and non-thermal X-ray emission, via resonant 
inverse Compton scattering of photons from the surface by the charged particles (e.g. \citealt{turolla15}). The resonant Compton scattering can manifest 
in the form of distinct soft and hard X-ray emission components. The former is fed by mildly relativistic charges, the latter by ultra-relativistic charges moving 
along an extended ($\lesssim$\,10 stellar radii) closed twisted magnetic loop around the star. The up-scattered photons convert to electron-positron pairs 
close to the star surface. The radiated energy is processed into a relativistic outflow of pairs that experiences a radiative drag as it propagates away from 
the star, and annihilates at the top of the loop. 
A dominant contribution from long-lived coronal currents to the emitted radiation can account naturally for the shrinking of the emitting region and the relatively 
high temperatures observed in \aa\ along the outburst decay (and seen also in several other magnetars; see e.g. \citealt{alford16,cotizelati17,borghese18}). 
It can also explain adequately a number of broad-band spectral and timing properties observed for this source over the past few years:

\emph{(i)} the emergent spectra of the resonant Compton scattering for mono-energetic uncooled relativistic electrons moving along field loops have 
been computed by \cite{wadiasingh18}. The slope of the PL tail at high energies is determined by the electron Lorentz factor, the loop orientation 
and the scattering kinematics. It strongly depends on the viewing angle and the location of resonant scattering, and is thus predicted to vary as a function 
of the rotational phase. This predicted behavior is consistent with our detection of a marked change in the values for the spectral parameters 
at the maximum and the minimum of the pulse profile in \aa;

\emph{(ii)} the hard X-ray emission is expected to be beamed along the magnetic loop, leading to detectable pulsed emission. The pulse profile shape strongly 
depends on the complex geometry of the emission from the electron-positron outflow, as well as the relative contribution of possible multiple loops to the observed 
emission \citep{wadiasingh18}. The loop does not necessarily extend above the small thermally-emitting hot spot on the star surface, along the line of sight. This 
is consistent with our detection of a misalignment between the peaks in the soft and hard X-ray pulse profiles of \aa. Photons are presumably up-scattered at 
different locations along the extended loop (i.e., at different heights above the stellar surface). Consequently, the hard X-ray emission shall be less anisotropic than 
the soft X-ray emission from the hot spot on the star surface. This is consistent with the smaller values for the pulsed fraction at higher energies in \aa. 

Hence, while both crustal cooling and coronal activity might be possibly at work in other magnetars, the properties observed for \aa\ seem to favour the latter 
scenario. However, besides this interpretative dichotomy (which is applicable to several other magnetars), the flux evolution of \aa\ gives some additional new 
clues, as discussed in the next section.

\subsection{A new persistent coronal state?}

Our analysis suggests that \aa\ is actually not decaying in luminosity any longer and has reached a persistent magnetospheric state that is different from the state 
observed in 2006, when the source was found at a much smaller luminosity. Such an interpretation is corroborated by the remarkable difference between the
values for the pulsed fraction measured at low energies in the past few years and in 2006.

A magnetospheric reconnection, most likely triggered by a sudden event in the crust (a thermoplastic wave or a starquake) can lead to a local reorganization of 
the magnetic field, leaving in the wake of the transient event a new pattern of twisted field associated with current bundles that close within the envelope and the crust. 
\cite{carrasco19} performed 3D relativistic force-free simulations of the magnetospheric outburst due to the progressive twisting of the line footprints 
at the surface, extending previous axisymmetric non-relativistic simulations \citep{parfrey13}. When a critical twist is reached, the magnetosphere undergoes 
a state of instability. This, in turn, leads to a rapid expansion of the lines, which finally reconnect, expelling plasmoids. After this rapid (millisecond-long) phase, a 
new state is reached, with non-negligible currents threading the magnetosphere.

The key issue is then to assess the survival timescale of the magnetospheric currents. A phenomenological estimate can be obtained simply by assuming that the 
measured luminosity ultimately comes from Ohmic dissipation of the magnetic twist: $ t_{{\rm state}}$\,$\sim$\,$E_{{\rm twist}}/L_{\rm X}$, where $E_{{\rm twist}}$ is 
the free magnetic energy contained in the twisted bundle. According to 3D \citep{carrasco19} and 2D \citep{beloborodov09,parfrey13} simulations, 
this magnetic energy is typically a fraction $f_{{\rm free}}$\,$\sim$\,0.01--0.1 of the total energy budget, so that $E_{{\rm twist}}$\,$\sim$\,$f_{{\rm free}}\,B_p^2R_\star^3/12$\,
$\sim$\,$10^{45}$\,$f_{{\rm free}}B_{14}^2R_{10}^3$\,erg, with $B_{14}$\,$=$\,$B_p/10^{14}$\,G and $R_{10}$\,$=$\,$R_\star/10$\,km. Thus we get 
\begin{equation}
\label{eq:tstate}
    t_{{\rm state}} \lesssim ~30~ \frac{f_{{\rm free}}}{0.1}~\frac{B_{14}^2R_{10}^3}{L_{35}} ~{\rm yr},
\end{equation}
where $L_{35}$ is the thermal luminosity in units of $10^{35}$ \lum.
For a large enough free magnetic energy, $E_{{\rm twist}}$\,$\propto$\,$f_{{\rm free}}B_p^2$ (i.e., a large twist and or a large background field), bundles could dissipate 
in tens of years, thus providing a persistent emission that is apparently constant on the typical observation timescales. On the other hand, smaller twists (associated 
with a smaller free magnetic energy) should dissipate faster. The estimate provided by Eq.\,(\ref{eq:tstate}) represents an upper limit, since it implicitly assumes a 
perfect efficiency in the Ohmic to radiative blackbody surface radiation. The more the Ohmic dissipation along the circuit is localized close to the surface, the more 
the approximation is correct.

A different, first-principles assessment of the survival timescales is provided by \cite{beloborodov09}. Using the estimate by \cite{beloborodov07} for 
the electrical potential along a field line due to  pair production and electrical discharge, he infers values in the range from years to decades (depending on 
the uncertain input parameters). These estimates were obtained from an axially symmetric calculation, assuming that the magnetosphere is globally twisted and that 
the star, where the current loops close, is perfectly conductive. However, the latter assumption is not accurate when considering the external envelope. As a matter 
of fact, preliminary estimates have shown that the very high resistivity of the outermost layers (with depths of meters or tens of meters) can account for the prolonged 
high temperatures observed for magnetars, if one supposes that most of the Ohmic dissipation of the magnetospheric currents occurs there \citep{akgun18,carrasco19}. 
These approximate estimates should then be fine-tuned by a better understanding and a more detailed computation of the total impedence along the current loops 
(dominated by the contribution from the magnetosphere and the envelope). Given their lack and the considerable uncertainties, the predicted timescales for 
the dissipation are compatible with the typical time span over which magnetars have been observed to linger at roughly the same flux (about a decade or so).

\section{Conclusions}
\label{sec:conclusions}

We studied the long-term evolution of the X-ray emission properties of the magnetar  \aa\ since February 2004, reanalyzing its three outbursts using new and 
archival observations with \swift, \nustar, \cxo\ and \xmm. About one year after the last outburst in 2009, the source settled on a relatively steady high flux level, 
a factor of $\sim$\,30 larger than its quiescent flux in 2006. Recent \nustar\ observations revealed faint hard X-ray emission up to about 70\,keV, and showed a 
peculiar progressive flattening of the high-energy power law spectral component in time. This is at variance with the cooling trend observed for the soft X-ray 
component, as well as with the typical overall gradual softening observed in other magnetar outbursts (see e.g. \citealt{esposito18}).

\aa\ provides compelling evidence that the persistent X-ray luminosity of magnetars should be necessarily dominated by magnetospheric currents, which
are responsible for both surface heating (in the form of thermal emission) and resonant Compton scattering (non-thermal soft and hard X-ray emissions). 
In fact, peculiar signatures for such a dominant magnetospheric component would be: ({\it i}) frequent changes between different persistent luminosities 
(not viable with internal crustal cooling alone); ({\it ii}) comparatively large persistent temperatures; ({\it iii}) small -- and possibly shrinking -- size of the emitting 
regions. Currently, not all magnetar outbursts have been followed over a long enough timescale to claim the full recovery to the pre-outburst level \citep{cotizelati18}. 
However, except for \aa, none of the sources monitored so far have shown such a large difference between their post-outburst steady flux and their previous 
quiescent level.

In the near future, an on-going monitoring campaign with \nicer\ (PI: A. Borghese; Borghese et al. in prep.) might reveal any peculiar timing behaviour 
possibly connected with the new state reached by \aa. Radio observations will also be valuable to reveal possible magnetospheric changes. Future \swift\ 
observations will play a crucial role to probe the flux steadiness of \aa\ over an even more extended time span and further characterize the emission properties 
of a magnetar that, at present, has no equal within this class of sources.

\begin{acknowledgements}
We thank the referee for helpful and valuable comments. FCZ acknowledges Matthew Baring for helpful discussions. FCZ, AB, NR and DV acknowledge support 
from grants SGR2017-1383 and PGC2018-095512-B-I00 and the support of the PHAROS COST Action (CA16214). FCZ is also supported by a Juan de la 
Cierva fellowship. NR, AB and DV are also supported by the ERC Consolidator Grant "MAGNESIA" (nr. 817661). DV acknowledges support from grants AYA2016-80289-P 
and AYA2017-82089-ERC. TE is supported by JSPS KAKENHI grant numbers 16H02198, 18H01246 and 18H04584. JAP acknowledges support from the Spanish 
Agencia Estatal de Investigaci\'on (grant PGC2018-095984-B-I00) and the Generalitat Valenciana (grant PROMETEO/2019/071).
The scientific results reported in this article are based on data obtained with \swift, the {\em Chandra X-ray Observatory}, \xmm\ and the \nustar\ mission. \xmm\ is 
an ESA science mission with instruments and contributions directly funded by ESA Member States and the National Aeronautics and Space Administration (NASA). 
The \nustar\ mission is a project led by the California Insitute of Technology, managed by the Jet Propulsion Laboratory, and funded by NASA. We made use of data 
supplied by the UK \swift\ Science Data Centre at the University of Leicester and of the XRT Data Analysis Software (XRTDAS) developed under the responsibility 
of the ASI Science Data Center (ASDC), Italy. 
We also made use of the \nustar\ Data Analysis Software (\textsc{nustardas} jointly developed by the ASI Science Data Center (ASDC, Italy) and the California Institute 
of Technology (USA), and of softwares and tools provided by the High Energy Astrophysics Science Archive Research Center (HEASARC), which is a service 
of the Astrophysics Science Division at NASA/GSFC and the High Energy Astrophysics Division of the Smithsonian Astrophysical Observatory. 
\end{acknowledgements}

\bibliography{biblio}{}
\bibliographystyle{aa_url}

\end{document}